\documentclass[a4paper]{article}

\usepackage{INTERSPEECH2022, comment}

\def\x{{\mathbf x}}
\def\y{{\mathbf y}}
\def\LM{\mathrm{LM}}
\def\ILM{\mathrm{ILM}}
\def\E2E{\mathrm{E2E}}
\def\AM{\mathrm{AM}}
\def\DR{\mathrm{DR}}
\def\res{\mathrm{res}}
\def\Score{\mathrm{Score}}
\def\T{\mathrm{T}}
\def\S{\mathrm{S}}
\def\L{\mathcal{L}}
\def\D{\mathcal{D}}

\title{Residual Language Model for End-to-end Speech Recognition}
\name{Emiru Tsunoo$^1$ Yosuke Kashiwagi$^1$ Chaitanya Narisetty$^2$ Shinji Watanabe$^2$}
\address{
  $^1$ Sony Group Corporation, Japan \\
  $^2$ Carnegie Mellon University, USA}
\email{emiru.tsunoo@sony.com}

\begin{document}

\maketitle
\begin{abstract}
End-to-end automatic speech recognition suffers from adaptation to unknown target domain speech despite being trained with a large amount of paired audio--text data.
Recent studies estimate a linguistic bias of the model as the internal language model (LM).
To effectively adapt to the target domain, the internal LM is subtracted from the posterior during inference and fused with an external target-domain LM.
However, this fusion complicates the inference and the estimation of the internal LM may not always be accurate.
In this paper, we propose a simple external LM fusion method for domain adaptation, which considers the internal LM estimation in its training. 
We directly model the residual factor of the external and internal LMs, namely the residual LM.
To stably train the residual LM, we propose smoothing the estimated internal LM and optimizing it with a combination of cross-entropy and mean-squared-error losses, which consider the statistical behaviors of the internal LM in the target domain data.
We experimentally confirmed that the proposed residual LM performs better than the internal LM estimation in most of the cross-domain and intra-domain scenarios.
%In Librispeech to TEDLIUM3 adaptation our method outperformed by a 4.0\% word-error-rate relative improvement with faster inference over the baseline shallow fusion.
%The results also indicated that the residual LM performs robustly in the intra-domain scenarios.
\end{abstract}

\section{Introduction}
End-to-end (E2E) automatic speech recognition (ASR) has attracted interest as a method of directly integrating acoustic models (AMs) and language models (LMs) because of its simple training and efficient decoding procedures. 
In recent years, various approaches have been studied, including connectionist temporal classification (CTC) \cite{graves06, %graves14, 
miao15, amodei16}, attention-based encoder--decoder models \cite{chorowski15, chan16}, hybrid models \cite{watanabe17,karita19}, and transducers \cite{
graves13rnnt,gulati2020,zhang2020transformer}.

E2E ASR requires pairs of audio and text data for training.
Even with a large amount of paired data, Del Rio et al. demonstrated that training with 960 h of Librispeech read speech does not result in sufficient performance in the mismatched domain of earnings calls \cite{delrio21_interspeech}.
If the target domain has paired data, adaptation techniques can be adopted \cite{yao12,delcroix16,Klejch18,tsunoo19_interspeech}.
However, in most scenarios, orders-of-magnitude more text-only data of the target domain are available, and it is more efficient to shift the linguistic bias of the E2E ASR model towards the domain of interest using such data.

%Although E2E ASR models cannot be explicitly factorized into AMs and LMs, m
Many researchers have studied fusion methods using an external LM trained with text-only data.
Shallow fusion, which linearly interpolates the E2E ASR model with an external LM, is the most popular approach \cite{chorowski15, hannun2014deep, kannan18}.
More structural integration can be observed in deep fusion \cite{gulcehre2015using}, cold fusion \cite{sriram18_interspeech}, and component fusion \cite{shan2019component}, which require additional training.
Fundamentally, the probability estimation of the LMs relies on softmax computation.
Therefore, although there are efficient log-sum-exp calculation tricks, it incurs a higher computational cost as the vocabulary size increases.
The density ratio approach \cite{mcdermott2019density} focuses more on domain adaptation by assuming that the source and target domains are acoustically consistent, and it adapts the E2E ASR model with LMs trained in each domain by following Bayes' rule.
Recently, the estimation of an internal LM, a linguistic bias of E2E ASR, has been investigated, and by subtracting from the ASR posterior, it improves performance in both cross-domain and intra-domain scenarios \cite{meng2021ilme,meng2021internal,zeineldeen2021investigating,zeyer2021librispeech}. 
However, both the density ratio and internal LM estimation complicate the inference computation.
In addition, due to the domain mismatch, the estimation of the internal LM may not always be accurate.

In this paper, we propose a simple external LM fusion for domain adaptation, which considers the internal LM.
Instead of subtracting the estimated internal LM and fusing with an external target-domain LM, we directly model the residual factor of them.
The difference of the probability distributions, namely the residual LM, is trained with a target-domain text-only dataset, considering the estimated internal LM in the specific domain.
Thus, the residual LM not only conveys the linguistic characteristic of the dataset, but also aggregates the estimation results of the internal LM in the target-domain corpus into the model, thereby alleviating the domain mismatch problem. 
In addition, because the distribution is no longer a probability, the residual LM can omit costly softmax computations in the output layer.
%While the residual LM is trained, the internal LM is estimated in the text-only target domain corpus; thus it conveys not only grammatical characteristics of the target corpus but also statistical behaviors of internal LM of given pre-trained E2E ASR model.
%We propose a combination of several optimization approaches to train such models.
We propose a training approach that applies smoothing to the internal LM probability, and combines cross-entropy with mean squared errors (MSEs) for the loss function.
%The trained residual LM can be fused in the same manner as shallow fusion, which simplifies the inference procedure by simultaneously considering the internal LM.
The trained residual LM can be simply fused in the same manner as shallow fusion.
We performed experiments to determine the effectiveness of the proposed residual LM in cross-domain and intra-domain scenarios using various corpora.
The results show that the proposed residual LM improves performance in cross-domain scenarios by 4.0\% relative word error rate (WER) in the Librispeech--TEDLIUM3 adaptation, with faster inference.
Additionally, the residual LM fusion method performs robustly in intra-domain scenarios.

\section{Formulation and Related Studies}
ASR is a problem in determining the most probable token sequence $\y$ given an input audio $\x$.
In a scenario in which the training and target domains differ, and the text corpus in the target domain is easily accessible, it is useful to combine the ASR model with an external LM trained using the target corpus.
With a Bayesian interpretation, classical hybrid ASR systems determine the highest probability by combining with the external LM, as follows:
\begin{align}
    p(\y|\x) = p(\x|\y;\theta_{\AM}) \cdot p(\y;\theta_\LM)\cdot \frac{1}{p(\x)}, \label{eq:hybrid}
\end{align} 
where $\theta_\AM$ and $\theta_\LM$ denote parameters of the AM and LM, respectively.
In E2E systems, $p(\x|\y;\theta_{\AM})$ is replaced by E2E neural networks, which can be further decomposed into the following terms using the Bayesian theorem:
\begin{align}
    p(\y|\x) = \frac{p(\y|\x;\theta_\E2E^\S)}{p(\y;\theta_\E2E^\S)}\cdot p(\x;\theta_\E2E^\S)\cdot p(\y;\theta_\LM^\T)\cdot \frac{1}{p(\x)}, \label{eq:asr}
\end{align}   
where $\theta_\E2E$ is a parameter set of an E2E ASR model. 
For domain adaptation, the E2E models are trained in a source domain, and the external LMs are trained in a target domain; thus, $*^\S$ and $*^\T$ represent source and target domains, respectively.
By omitting $p(\x)$ and $p(\x;\theta_\E2E^\S)$, which are not required to search for the highest probability of $\y$, the score function for recognition can be expressed in a logarithmic scale as
\begin{align}
    \Score(\y|\x) = \log p(\y|\x;\theta_\E2E^\S) &- \log p(\y;\theta_\E2E^\S) \nonumber \\
    &+ \log p(\y;\theta_\LM^\T). \label{eq:score}
\end{align}
The first term is the output posterior of the E2E ASR neural network, and the second term is the implicit linguistic bias (prior) of the trained E2E model.
The last term is an external LM trained in the target-domain text corpus.

\subsection{Shallow fusion}
E2E ASR models are often used with an external LM trained with a text-only corpus of the target domain. 
% This is reasonable because, unlike E2E ASR training, which requires audio--text pair data, external LM training requires only text data, which can be easily obtained with a large scale.
This is reasonable because, unlike the E2E ASR requiring a large corpus of paired audio--text data for training, external LM training requires only text data, which can be easily obtained with a large scale.
Shallow fusion is a common method for integrating an E2E ASR model with an LM \cite{chorowski15, hannun2014deep, kannan18}.
In practice, the second term in Eq.~(\ref{eq:score}) is omitted because it is intractable, and in shallow fusion, only the external LM term is combined by introducing an LM weight, $\lambda_\LM$ as:
\begin{align}
    \Score(\y|\x) = \log p(\y|\x;\theta_\E2E^\S) + \lambda_{\LM} \log p(\y;\theta_\LM^\T). \label{eq:sf}
\end{align}

\subsection{Density ratio approach}
The density ratio \cite{mcdermott2019density} assumes that the source and target domains are acoustically consistent. %the AM and LM factors in a source-domain E2E model, and the source-domain and target-domain AMs are invariant.
%Based on Bayes' rule, it calculates the ratio between the probabilities of the source- and target-domain LM in decoding.
%By applying Bayes' rule, it calculates the ratio between the source- and target-domain linguistic priors, and applies it to the source-domain posterior to estimate the target-domain posterior.
Assuming that an LM trained with text-only data in the source domain can represent the linguistic prior of the E2E model, i.e., the second term of Eq.~(\ref{eq:score}), the density ratio uses the ratio of LMs trained in both domains for adaptation.
Thus, the score in the density ratio approach can be expressed using the source-domain and target-domain LM weights, $\lambda_\DR$ and $\lambda_\LM$, as follows:
\begin{align}
    \Score(\y|\x) = \log p(\y|\x; \theta_\E2E^\S) &- \lambda_\DR \log p(\y;\theta_\LM^S) \nonumber \\ &+ \lambda_\LM \log p (\y;\theta_\LM^T). \label{eq:dr}
\end{align}

\subsection{Internal language model estimation}
\label{ssec:ilme}
Internal LM estimation (ILME) \cite{meng2021ilme, meng2021internal, zeineldeen2021investigating, zeyer2021librispeech} attempts to estimate the second term in Eq.~(\ref{eq:score}), i.e., $\log p(\y;\theta_\E2E^\S)$.
As a result, the score for decoding using the internal LM is expressed as follows.
\begin{align}
    \Score(\y|\x) = \log p(\y|\x; \theta_\E2E^\S) &- \lambda_\ILM \log p(\y;\theta_\E2E^S) \nonumber \\ &+ \lambda_\LM \log p (\y;\theta_\LM^T), \label{eq:ilm}
\end{align}
where $\lambda_\ILM$ is a weight parameter for the estimated internal LM.

A common method of estimating the internal LM, $p(\y;\theta_\E2E^S)$, is to replace the encoder output with zero-filled vectors and infer only with the decoder.
This is because, particularly in attention-based encoder--decoder and transducer architectures, decoders function similarly to LMs as they estimate next $i$-th token $y_i$ with a given previous output $\y_{1:i-1}$.
This estimation approach is statistically reasonable in the source domain because the encoder output is most likely normalized to zero-mean vectors with a normalization layer.
%However, it might not be the case in the different target domain and the behavior of estimated internal LM is unpredictable because the E2E ASR model is trained only with the source domain dataset.
%However, this might not be possible if there is a domain mismatch, and the behavior of the internal LM becomes unpredictable when it is estimated with the mismatched domain speech.
However, this estimation is performed in every inference, which complicates the computation.
In addition, when it is estimated with the mismatched domain speech, the behavior of the internal LM becomes unpredictable and the estimation may not always be accurate.

\section{Residual Language Model}
\subsection{Definition of the residual language model}
\label{ssec:definition}
%Instead of estimating the internal LM in every inference, we propose to directly model the target-domain external LM from which the estimated internal LM of the source-domain is already subtracted.
Instead of estimating the internal LM in every inference, we propose to directly model the residual factor of the target-domain external LM and the estimated internal LM.
The model predicts the difference between the second and third terms (internal and external LMs, respectively) in Eq.~(\ref{eq:score}), which we defined as the residual LM.
By directly modeling the residual term, we can simplify the inference computation to shallow fusion, as follows:
\begin{align}
    \Score(\y|\x) = \log p(\y|\x; \theta_\E2E^\S) + \lambda_\LM f(\y;\theta_{\res}^T), \label{eq:res}
\end{align}
where $f(\y)$ is the proposed residual LM.
The residual LM conveys both the second and third terms in Eq.~(\ref{eq:score}); thus Eq.~(\ref{eq:res}) strictly follows the score calculation (\ref{eq:score}) derived by Bayes' interpretation.
The main differences from ILME are as follows.
\begin{itemize}
    \item The residual LM models the residual factor of an external LM and the internal LM, which simplify the inference procedure and reduce computational cost.
    \item The difference of two LMs is no longer a probability distribution; thus it can further omit the log-softmax operation, which requires costly log-sum-exp calculation.
    \item The residual LM conveys statistical behavior of the estimated internal LM in the target-domain text corpus.
\end{itemize}

The residual LM is trained using the target-domain text-only data.
To model the residual terms of the external and internal LMs, we can define the training target of the model output, $r(\y)$ as follows:
\begin{align}
    %r(\y) = \frac{q^*}{p(\y;\theta_{\E2E}^S)^\gamma}
    r(\y) = \log q_\y^* - \gamma \log p(\y;\theta_{\E2E}^S),\label{eq:target}
\end{align}
where $q_\y^*$ is a reference label, and $\gamma$ is a tunable parameter.
To avoid log-zero computation, a smoothed label as in \cite{szegedy2016rethinking} is adopted for the reference $q_\y^*$, as 
\begin{align}
    q^*_{y,k} = (1-\omega)\delta_{y,k} + \frac{\omega}{K},
\end{align}
where $k$ is the vocabulary index, $\delta_{y,k}$ is Dirac delta with respect to $k$, $K$ is the vocabulary size, and $\omega$ is the smoothing weight.
Throughout the training data, the internal LM is estimated using Eq.~(\ref{eq:target}).
While ordinary ILME (Sec.~\ref{ssec:ilme}) is performed only with the input speech sample, the residual LM statistically considers its behavior in the entire target-domain training data, which may be effective during inference with the target-domain data.

\subsection{Smoothing of the internal language model}
\label{ssec:smoothing}
Because the estimation of the internal LM is not always reliably preformed in the target domain, the probability distribution $p(\y;\theta_{\E2E}^S)$ may be inaccurate.
If the probability of the token of interest is incorrectly estimated to be significantly low, the target distribution $r(\y)$ diverges to infinity.
To prevent this, we use a temperature $T>1$, as introduced in \cite{hinton2015distilling}, to soften its distribution.
\begin{align}
    \tilde{p}(\y;\theta_{\E2E}^S)_i = \frac{\exp(z_i/T)}{\sum_{i=1}^{K}\exp(z_i/T)} + \epsilon, \label{eq:temperature}
\end{align}
where $z_i$ is the output of the decoder of the E2E ASR model.
We further introduce a small value, $\epsilon$ to avoid log zero in Eq.~(\ref{eq:target}).
By replacing $p(\y)$ with $\tilde{p}(\y)$, the softened target is used instead as
\begin{align}
    %\tilde{r}(\y) = \frac{q^*}{\tilde{p}(\y;\theta_{\E2E}^S)^\gamma}.
    \tilde{r}(\y) = \log q_\y^* - \gamma \log\tilde{p}(\y;\theta_{\E2E}^S).\label{eq:smooth_target}
\end{align}

\subsection{Training of the residual language model}
\label{ssec:training}
The residual LM is trained to minimize the distance between the target distribution $\tilde{r}(\y)$ and the model output $f(\y)$.
A straightforward approach is to minimize the L1 norm or MSEs between the model output and the target distribution.
However, in our preliminary experiments, the trained model did not reasonably perform.
We assume that this was because the Euclidean-based optimization attempted to minimize the distances of all vocabulary entries equally, which did not contribute to improving recognition performance.

To stably train the residual LM, we decompose the target function by introducing a normalization term, as follows:
\begin{align}
    \tilde{r}(\y) &= \log q(\y) + \log N(\y), \label{eq:factor}
\end{align}
where $q(\y)$ and $N(\y)$ are defined as 
\begin{align}
    q(\y) &= \mathrm{softmax}\left(\log q_\y^* - \gamma \log\tilde{p}(\y;\theta_{\E2E}^S)\right) \nonumber \\
    &= \frac{q_\y^*}{\tilde{p}(\y;\theta_{\E2E}^S)^\gamma}\cdot \frac{1}{N(\y)}, \\
    N(\y)&=\sum_{k=1}^{K}\frac{q_{\y,k}^*}{\tilde{p}(\y;\theta_{\E2E}^S)_k^\gamma}. \label{eq:bias}
\end{align}
Thus, the target function $\tilde{r}(\y)$ is decomposed into the probabilistic term, $\log q(\y)$, and the bias term, $\log N(\y)$.
We propose to separately minimize the distances pertaining to these terms, with a combination of cross-entropy of probabilities and the Euclidean distance of the biases. 

\subsubsection{Cross-entropy loss for the probabilistic term}
\label{sssec:ce}
LMs are generally optimized by minimizing the negative log likelihood by computing the cross entropy, which is equivalent to minimizing the perplexity.
Inspired by this, we apply cross-entropy to optimize the probabilistic term, $\log q(\y)$, in Eq.~(\ref{eq:factor}).
To extract the probabilistic factor from the residual LM, we marginalize the model output by applying softmax function.
\begin{align}
        p(\y;\theta_{\res}^T) = \mathrm{softmax}(f(\y;\theta{\res}^T)).
\end{align}
Then a cross-entropy loss is accumulated in the target-domain dataset $\D_T$ as
\begin{align}
    \L_{\mathrm{CE}} = -\sum_{\y\in \D_{T}}\sum_i q(y_i)\log p(y_i;\theta{\res}^T). \label{eq:ce}
\end{align}

\subsubsection{Mean-squared-error loss for the bias term}
\label{sssec:mse}
We also aim to minimize the distance between bias terms of the target and model.
%In practice, this depends on the context of $\y$, i.e., $\y_{0:i-1}$ for the current token index $i$, because the internal LM $p(\y;\theta_\E2E^S)=\prod_{i=1}^I p(y_i|y_{0:i-1};\theta_\E2E^S)$ depends on the context.
The bias of the residual LM can be derived as 
\begin{align}
    f(y;\theta_\res^T)-\log p(y;\theta_\res^T) = \log \sum_{k=1}^K \exp(f(y_k;\theta_\res^T)).
\end{align}
The bias term of the target, $\log N(\y)$ in Eq.~(\ref{eq:factor}), is also a scalar value, % which is
defined in Eq.~(\ref{eq:bias}).

We further simplify those terms to define an objective function.
In the case that the label $q_{y_i}^*$ is a hard label, i.e., $\omega = 0$, Eq.~(\ref{eq:bias}) becomes
\begin{align}
    %N(y_i)=\frac{1}{p(y_i|\y_{0:i-1};\theta_\E2E^S)^\gamma}. \label{eq:app_bias}
    N(y_i)=\frac{1}{\tilde{p}(y_i;\theta_\E2E^S)^\gamma}. \label{eq:app_bias}
\end{align}
Therefore, we approximate the bias using Eq.~(\ref{eq:app_bias}) and define an MSE loss as
\begin{align}
    \L_{\mathrm{MSE}} = \frac{1}{|\D_{T}|}\sum_{\y\in \D_{T}}\sum_{i}(f(y_i;\theta_\res^T)&-\log p(y_i;\theta_\res^T) \nonumber \\ 
    &+\gamma\log \tilde{p}(y_i;\theta_\E2E^S))^2. \label{eq:mse}
\end{align}

\subsubsection{Integrated objective function}
\label{sssec:objective}
%Finally, we propose to optimize the model with a hybrid manner of aforementioned optimization.
Finally, we integrate aforementioned optimization in a hybrid manner.
The cross-entropy and MSE losses are combined with a weighted sum using a parameter $\eta$ as follows.
\begin{align}
    \L = \L_{\mathrm{CE}} + \eta\L_{\mathrm{MSE}}. \label{eq:integ}
\end{align}
Note that both losses contain statistic terms of the internal LM, in $q(y_i)$ of Eq.~(\ref{eq:ce}) and in Eq.~(\ref{eq:mse}).
Therefore, the trained residual LM considers the statistical behaviors of the internal LM in the target-domain data.

\section{Experiments}

\subsection{Cross-domain evaluation}
\label{ssec:cross}

To evaluate the effectiveness of the proposed residual LM on domain adaptation, we evaluated cross-domain scenarios in English and Japanese.
\subsubsection{Experimental setup}
\label{sssec:setup}
For the English evaluation, we trained an E2E ASR model with the Librispeech dataset \cite{panayotov15}, a read speech corpus, and applied it to the TED-LIUM 3 \cite{hernandez2018ted} dev/test set, which is a spontaneous lecture style.
For Japanese, we trained an E2E ASR model with the lecture style CSJ corpus \cite{csj}, and then applied it to the LaboroTV dev set \cite{ando2021construction}, a corpus of TV programs.
We trained streaming Transformer E2E ASR models following \cite{tsunoo2022run}.
%using the English LibriSpeech dataset \cite{panayotov15}, AISHELL-1 \cite{aishell17} Mandarin tasks, and the Japanese CSJ dataset \cite{csj}.
%Those were considered as source domain dataset.
The input acoustic features were 80-dimensional filter bank features.
The transformer architecture consisted of 12 encoder blocks and six decoder blocks, with four-head 256-unit attention layers and 2048-unit feed-forward layers.
Contextual block encoding \cite{tsunoo19} was applied to the encoder with a block size of 40, a shift size of 16, and a look-ahead size of 8.
The models were trained using multitask learning with CTC loss, as in \cite{watanabe17}, with a weight of 0.3.
We used the Adam optimizer and Noam learning rate decay, and applied SpecAugment \cite{park19}.

External LMs for baseline shallow fusion \cite{hannun2014deep} as well as the proposed residual LMs were trained using the text-only data of the training set in the target corpora, i.e. TED-LIUM3 and LaboroTV.
Both LMs were four-layer unidirectional LSTM with 1024 units for the English task and two-layer unidirectional LSTM with 2048 units for Japanese.
We applied the byte-pair encoding subword tokenization with 5000 token classes for English LMs.
The tokens for Japanese LMs had 3262 character classes.
The training weight in Eq.~(\ref{eq:target}) was set as $\gamma=0.3$.
We set the temperature in Eq.~(\ref{eq:temperature}) as $T=2$ and the loss integration weight in Eq.~(\ref{eq:integ}) as $\eta=0.1$ for all experiments.

In addition to shallow fusion, we compared residual LMs with density ratio \cite{mcdermott2019density} and ILME \cite{meng2021ilme}.
%Both the residual LMs trained with MSE optimization (Sec.~\ref{ssec:mse}) and KL-divergence optimization (Sec.~\ref{ssec:kl}) were evaluated.
After the parameter search, the LM weight was set to $\lambda_\LM=0.6$, and the weight for density ratio in Eq.~(\ref{eq:dr}) was $\lambda_\DR=\{0.1,0.3\}$, for the respective dataset.
The internal LM weight in Eq.~(\ref{eq:ilm}) was set to $\lambda_\ILM=0.3$ for both languages.
The beam size for decoding was 10.
The internal LM was estimated by replacing the encoder output with a zero-filled tensor as in \cite{meng2021ilme, zeineldeen2021investigating}.
We also measured the inference speed using randomly sampled 100 utterances from each dev set of the target domain.

\begin{comment}
\begin{table*}[t]
  \caption{Librispeech WERs in an intra-domain scenario.}
  \label{tab:librispeech}
  %\vspace{1mm}
  \vspace{-0.3cm}
  \centering
  \scalebox{0.9}{
  \begin{tabular}{l|cccc|c}
    \hline
    %&\multicolumn{5}{c}{Librispeech} \\
     %&\multicolumn{4}{c|}{(WER)} & decoding  \\
     & dev-clean & dev-other & test-clean & test-other & Decoding speed\\
    \hline\hline
    % Shallow Fusion \cite{hannun2014deep} & 2.7 & 7.2 & 2.9 & 7.4  \\
    Shallow Fusion \cite{hannun2014deep} & 1.9 & 5.1 & 2.2 & 5.1  & x1.0 \\
    % ILME \cite{meng2021ilme} & {\bf 2.6} & 7.4 & {\bf 2.8} & 7.4  \\
    ILME \cite{meng2021ilme} & 1.9 & 5.1 & 2.2 & 5.1 & x0.51 \\
    \hline
    Residual LM w/ KL &&&&  \\
    % \ \ {\it w/o Normalization} (\ref{eq:temperature}) &&&&&{5.9}&{6.5} \\
    % \ \ {\it w/o Smoothing} (Sec.~\ref{ssec:smoothing}) &2.8&7.7&2.9&7.7\\
    % \ \ {\it w/o Both} (\ref{eq:wlm}) &&&&&{\bf 5.6} &6.2& {\bf 5.9} & {\bf 4.3} & {\bf 4.8}  \\
    
    \hline
  \end{tabular}
  }
  %\vspace{-0.4cm}
\end{table*}
\end{comment}

\begin{table}[t]
  \caption{ASR results in cross-domain adaptation scenarios.}
  \label{tab:crossdomain}
  %\vspace{1mm}
  \vspace{-0.3cm}
  %\centering
%  \begin{tabular}{l|cc|cc}
  \hspace{-0.5cm}
  \scalebox{0.9}{
  \begin{tabular}{l|ccc|cc}
    \hline
    &\multicolumn{3}{c|}{LS $\rightarrow$ TEDLIUM3} & \multicolumn{2}{c}{CSJ $\rightarrow$ LaboroTV} \\
    &\multicolumn{3}{c|}{(WER)} & \multicolumn{2}{c}{(CER)} \\
     & Dev  & Test & Dec. Speed &Dev & Dec. Speed\\
    \hline\hline
    % No LM & & & &27.1 & 1.54\\
    Shallow Fusion \cite{hannun2014deep} & 13.2 & 12.6 & x1.0 & 24.6 & x1.0 \\
    Density Ratio \cite{mcdermott2019density} & 12.9 & 12.7 &x0.92 & {\bf 21.9} &x0.97 \\
    ILME \cite{meng2021ilme} & {12.9} & {12.2}  & x0.58 & 23.7 & x0.58  \\
    \ \ {\it w/ Smoothing} & {12.9} & {12.2} &---& 24.3 & ---\\
    \hline
    %Residual LM w/ MSE (Sec.~\ref{ssec:mse}) & {\bf } & {\bf } & 24.1 \\
    %Residual LM w/ KL (Sec.~\ref{ssec:kl}) & {12.9} & {12.4} & {22.7} \\
    % Residual LM w/ KL (Temperature) & {12.9} & {12.4} & {22.7} \\
    % Residual LM w/ KL (AddEps) & {\bf 12.6} & {\bf 12.1} & {23.1} \\
    Residual LM & {\bf 12.6} & {\bf 12.1} & {\bf x1.08}& {22.7} & {\bf x1.04} \\
    % \ \ {\it w/o Normalization} (\ref{eq:temperature}) & & & {\bf 22.9} \\
    \ \ {\it w/o Smoothing} & {12.9} & {12.2} &---& 24.1 &---\\
    % \ \ {\it w/o Both} (\ref{eq:wlm}) & 18.2 & 17.7 & 23.2 \\
    %(Residual LM + Density Ratio) &  & & ({\bf 21.8}) \\
    
    \hline
  \end{tabular}
  }
  \vspace{-0.4cm}
\end{table}

\subsubsection{Experimental results}
\label{sssec:cross_results}
The experimental results are listed in Table.~\ref{tab:crossdomain}.
In both the English and Japanese scenarios, the density ratio approach and ILME achieved lower WERs than the baseline shallow fusion.
%The residual LM trained with MSE loss degraded performance from the Shallow Fusion.
%We assume that MSE loss minimized distance even in irrelevant vocabulary entry, which did not directly contribute to its performance.
%On the other hand, optimizing with KL-divergence successfully model the residual of LMs, with the normalized form in Eq.~(\ref{eq:norm}).
The proposed residual LM performed better than ILME and achieved the best performance in the English adaptation with a WER of 12.1 \% on the test set (4.0 \% WER relative improvement over the shallow fusion).
Although, for the Japanese task, the residual LM performed poorer than the density ratio approach, there was an improvement from the baseline shallow fusion and ILME. %, which is mathematically equivalent. 
We assume that the residual LM can consider the statistical behavior of the internal LM, as discussed in Sec.~\ref{ssec:definition}.

The results of the decoding speed are shown relative to shallow fusion as a base.
ILME required almost twice a decoding duration as shallow fusion, because the decoders of the E2E ASR models were required to compute twice, once for ASR and once for ILME.
%The density ratio approach consumed more time because it computed two external LSTM LMs, which are slower in inference than the transformer architectures of the E2E ASR decoders.
The density ratio was also slightly slower than the shallow fusion because it was required to compute two LMs.
The proposed residual LM was slightly faster than the baseline shallow fusion, because it can omit softmax operation, particularly in the larger vocabulary size in English setup.

We performed further ablation studies on the proposed residual LM. %with KL-divergence optimization.
When we replaced the smoothed target $\tilde{r}(\y)$ with the regular target $r(\y)$, as defined in Eq.~(\ref{eq:target}), we observed a significant performance drop from the smoothed target in the Japanese case.
We assume that the internal LM estimation is not always accurate in the target domain. %, and the token distribution in Japanese was more sparse than that in English.
On the other hand, applying smoothing did not aid the ILME to improve performance.
% This may be because smoothing was statistically effective in training the target distribution of the residual LM.
We assume that using the smoothed soft labels in training gains positive effect similarly to the knowledge distillation learning \cite{hinton2015}.

\subsection{Intra-domain evaluation}
\subsubsection{Chinese/Japanese corpora}
We performed an intra-domain evaluation to determine if the proposed residual LM did not have an adverse impact in the matched conditions.
The residual LMs were evaluated using AISHELL-1 \cite{aishell17} and in the CSJ evaluation set.
The experiments followed the configuration in Sec.~\ref{sssec:setup}, except
the LMs for AISHELL-1 consisted of two LSTM layers with 650 units, whose output is 4233 character classes. %, and for the other tasks, the model architecture followed the setup described in Sec.~\ref{sssec:setup}.
%We compared the results with the shallow fusion and ILME as the baselines.
The LM weights were set to $\lambda_\LM=\{0.2, 0.3\}$ in the respective evaluation sets, the weights for the ILME were $\lambda_\ILM=\{0.1, 0.3\}$, and the training parameters were $\gamma=\{0.1,0.3\}$ respectively.

The character error rate (CER) results are presented in Table.~\ref{tab:intradomain}.
In our reproduction of ILME, even with the best effort to search for the parameters, we observed degradation in both the AISHELL-1 and CSJ evaluation sets.
The literature reported that ILME had a positive effect even in the intra-domain scenarios \cite{meng2021ilme}, but they were evaluated only in the English dataset, and were not tested in other languages.
In contrast, our proposed residual LM performed robustly throughout the corpora and even had lower CERs particularly in the Japanese scenario.
%In addition, we observed that the smoothing in Sec.~\ref{ssec:smoothing} was not as effective as in the cross-domain experiments.
%We assume that this is because of the correctness of the estimation of the internal LMs in the matched domain test sets.

\subsubsection{Transformer LM evaluation using Librispeech}
Lastly, we evaluated the state-of-the-art architecture using the Librispeech dataset.
For the E2E ASR model, we adopted 12 conformer encoder blocks \cite{gulati2020} with 512-unit eight-head attention and 2048 unit feed forward layers, followed by six transformer decoder with 512-unit eight attention heads and 2048 unit feed forward layers.
The LMs were 16-layer transformers.
%The LM weight was $\lambda_\LM=0.6$, and we used the best parameter for the ILME; $\lambda_\ILM=0.3$.
We set the parameters as $\lambda_\LM=0.6$, $\lambda_\ILM=0.3$ and the beam size was 30.
Inference speed was also evaluated using the test-clean set.

Table.~\ref{tab:librispeech} shows the results.
No significant difference was observed between the shallow fusion and ILME in this setup.
Although we observed slight degradation in the test-other set, both ILME and the proposed residual LM performed robustly with the large model architecture.
We observed similar tendency of inference speed as in Table.~\ref{tab:crossdomain}.

\begin{table}[t]
  %\caption{Intra-domain CER results in Chinese/Japanese corpora.}
  \caption{Chinese/Japanese CERs in intra-domain scenarios.}
  \label{tab:intradomain}
  %\vspace{1mm}
  \vspace{-0.3cm}
  \centering
  \scalebox{0.9}{
  \begin{tabular}{l|cc|ccc}
    \hline
     & \multicolumn{2}{c|}{AISHELL-1} &\multicolumn{3}{c}{CSJ} \\
     %& \multicolumn{2}{c|}{(CER)} &\multicolumn{3}{c}{(CER)} \\
     &  Dev & Test & eval1 & eval2 & eval3 \\
    \hline\hline
    Shallow Fusion \cite{hannun2014deep} & 5.7 & 6.3 & 6.0 & 4.4 & 5.1 \\
    ILME \cite{meng2021ilme} &  5.8 & 6.4 & 6.3 & 4.5 & 5.3 \\
    \hline
    Residual LM &{\bf 5.6}&{\bf 6.1} & {\bf 5.5} & {\bf 4.2} & {\bf 4.6}  \\
    % \ \ {\it w/o Normalization} (\ref{eq:temperature}) &&&&&{5.9}&{6.5} \\
    %\ \ {\it w/o Smoothing} &{\bf 5.6}&{\bf 6.1} & {5.7} & {4.4} & {4.9} \\
    % \ \ {\it w/o Both} (\ref{eq:wlm}) &&&&&{\bf 5.6} &6.2& {\bf 5.9} & {\bf 4.3} & {\bf 4.8}  \\
    
    \hline
  \end{tabular}
  }
  %\vspace{-0.4cm}
  \vspace{-0.2cm}
\end{table}

\begin{table}[t]
  \caption{Librispeech WERs in an intra-domain scenario.}
  \label{tab:librispeech}
  %\vspace{1mm}
  \vspace{-0.3cm}
  \centering
  \scalebox{0.9}{
  \begin{tabular}{l|cc|c}
    \hline
    %&\multicolumn{5}{c}{Librispeech} \\
     %&\multicolumn{4}{c|}{(WER)} & decoding  \\
     & test-clean & test-other & Dec. speed\\
    \hline\hline
    % Shallow Fusion \cite{hannun2014deep} & 2.7 & 7.2 & 2.9 & 7.4  \\
    %Shallow Fusion \cite{hannun2014deep} & 2.2 & 5.1  & x1.0 \\
    Shallow Fusion \cite{hannun2014deep} & {\bf 2.3} & {\bf 5.2}  & x1.0 \\ %7epoch
    % ILME \cite{meng2021ilme} & {\bf 2.6} & 7.4 & {\bf 2.8} & 7.4  \\
    %ILME \cite{meng2021ilme} &  2.2 & 5.1 & x0.51 \\
    ILME \cite{meng2021ilme} &  {\bf 2.3} & {\bf 5.2} & x0.51 \\ %7epoch
    \hline
    Residual LM  &{\bf 2.3}& 5.3 &{\bf x1.13}\\ %7epoch
    % \ \ {\it w/o Normalization} (\ref{eq:temperature}) &&&&&{5.9}&{6.5} \\
    % \ \ {\it w/o Smoothing} (Sec.~\ref{ssec:smoothing}) &2.8&7.7&2.9&7.7\\
    % \ \ {\it w/o Both} (\ref{eq:wlm}) &&&&&{\bf 5.6} &6.2& {\bf 5.9} & {\bf 4.3} & {\bf 4.8}  \\
    
    \hline
  \end{tabular}
  }
  \vspace{-0.4cm}
  %\vspace{-0.4cm}
\end{table}
% softmaxを外す方がよいかも
% MSEでやるとき、1. L1でやる、2. 分母と分子を分けて扱う(たとえば分子はL1、分母はなんとか) 3. 重みつけてもよい
% smoothing二つ合わせても良い

\section{Conclusion}
We propose a simple external LM fusion method for domain adaptation, which considers the internal LM estimation in its training.
We directly modeled the ratio of an external target domain LM to an internal LM of the E2E ASR model, which is called residual LM.
The residual LM was stably trained using a combination of cross-entropy and MSE losses. %, considering the statistical behaviors of the internal LM in the target domain data.
The experimental results indicated that the proposed residual LM performed better than the internal LM estimation in most of the cross-domain and intra-domain scenarios.

%We experimentally confirmed that the proposed residual LM performs better in cross-domain scenarios.% by 4.0\% WER relative improvement with faster inference in the Librispeech to TEDLIUM3 adaptation, for instance.
%The results also demonstrated that the residual LM performed robustly in intra-domain scenarios.

\bibliographystyle{IEEEtran}

\bibliography{mybib}

\end{document}